\documentclass[preprint,floats,aps,epsfig,nofootinbib,axodraw,amssymb,11pt]{revtex4}

\usepackage{subfigure,epsfig}
\def\beq{\begin{equation}}
\def\eeq{\end{equation}}
\def\bea{\begin{eqnarray}}
\def\eea{\end{eqnarray}}

\def\etmiss{E\!\!\!\!\slash_{T}}
\def\ptmiss{p\!\!\!\slash_{T}}
\def\pslash{\not{\hbox{\kern-4pt $p$}}}
\def\qslash{\not{\hbox{\kern-4pt $q$}}}
\def\lv{\not{\hbox{\kern-4pt $L$}}}
\def\lsim{\mathrel{\raise.3ex\hbox{$<$\kern-.75em\lower1ex\hbox{$\sim$}}}}
\def\gsim{\mathrel{\raise.3ex\hbox{$>$\kern-.75em\lower1ex\hbox{$\sim$}}}}
\def\ifmath#1{\relax\ifmmode #1\else $#1$\fi}

\usepackage{graphicx}
\usepackage{bm}
\begin{document}
\draft
\renewcommand{\thefootnote}{\arabic{footnote}}

\begin{flushright}
MADPH-09-1549, NPAC-09-15 \\
\end{flushright}

\title{Genuine CP-odd Observables at the LHC}
\bigskip
\author{Tao Han and Yingchuan
Li\footnote{Email Address: than@hep.wisc.edu, \  yli@physics.wisc.edu}}
\address{Department of Physics, University of Wisconsin, 1150 University Avenue,
Madison, WI 53706, U.S.A.}

\begin{abstract}
We discuss how to construct genuine CP-odd observables at the LHC.
We classify the observables according to the even and odd properties
under the naive T-transformation ($\hat{T}$). There are two classes
of observables of our interests: CP-odd and $\hat{T}$-even; CP-odd
and $\hat{T}$-odd. We expect them to have broad applications to many
processes in theories beyond Standard Model with CP violation. For
the purpose of illustration, we use simple example of $W^+W^-$
production and subsequent decays at the LHC, where the CP violation
effects are parameterized by effective CP-violating operators of
$WWZ$ coupling. We find significant sensitivity to the CP-odd
couplings.
\end{abstract}

\maketitle

\vskip 0.1cm
\noindent

\section{Introduction}
The non-invariance of
fundamental interactions under the Charge-conjugate and parity (CP)
transformations still possesses a major challenge in physics. The
successful parameterization of CP violation \cite{Kobayashi:1973fv}
in the electroweak Standard Model (SM) fits the observations in the
kaon and $B$ systems  nearly perfectly \cite{Amsler:2008zzb}. Yet,
our mere existence,
namely, the matter dominance over anti-matter in our observed
Universe, can only be explained by new CP violation sources beyond SM.
Thus seeking for new effects of CP violation is always of
fundamental importance. CP violation can be non-ambiguously
identified only via CP-odd observables, that can be constructed
either by comparing two CP-conjugate processes $i \rightarrow f$ and
$\bar{i} \rightarrow \bar{f}$, or by observing transitions between
CP-odd and CP-even eigenstates.

The Large Hardron Collider (LHC) at CERN will lead us to revolutionary
discoveries of new physics beyond the SM, which quite generically introduces
new sources of CP violation. The importance of searching
for CP-violation effects at the LHC are thus strongly motivated,
and cannot be overemphasized.
However, it is quite challenging to construct genuine CP-odd
observables at the LHC. First, the initial state of the LHC as a $pp$
collider is not a CP eigenstate, in contrast to the neutrality property of an
$e^+e^-$ collider or a $p\bar p$ collider. One may have to seek for
subprocesses ($q\bar{q},\ gg$ initial states) or decays of CP
eigenstates at the LHC experiments.
Second, even with the initial states $q(p_1) \bar{q}(p_2)$ and
$g(p_1) g(p_2)$, they form a CP eigenstate only in their
center-of-mass (cm) frame which is in general different from the lab
frame. These two reference frames are related by a longitudinal
boost that is unknown and different event by event.
%
Furthermore,  the symmetric beams of  $pp$ at the LHC make it impossible to identify
the direction of a quark versus an anti-quark on an event-by-event basis,
which is often needed when making use of their momenta.

Assuming that CPT transformation is a good symmetry, CP-violation is
equivalent to T-violation. Due to the difficulty to directly
construct variables that are odd under T-transformation
\begin{eqnarray}
\nonumber
{\cal T} |\vec{p},\vec{s}>=<-\vec{p},-\vec{s}|,
\end{eqnarray}
attempts have been mainly made in the literature to construct
$\hat{T}$-odd observables, where $\hat{T}$ is the ``naive"
time-reversal transformation \cite{Valencia:1994zi}, only to change
the sign of momenta and spins,
\begin{eqnarray}
\nonumber
\hat{T} |\vec{p},\vec{s}>=|-\vec{p},-\vec{s}>,
\end{eqnarray}
 without reversing the initial and final states nor other internal quantum
numbers. While a  $\hat{T}$-odd observable may not be necessarily a
CP-odd observable, the transformation is useful to classify CP-odd
observables.  In general, there can be two types of phases in an
amplitude. One is the CP-violating phase $\theta$ from a fundamental
theory, and the other is CP-conserving phase $\delta$ coming from
the absorptive part of an amplitude when the intermediate states in
the loop become on-shell. General arguments  \cite{Valencia:1994zi}
show that  CP-odd observables can  be either $\hat{T}$-even or
$\hat{T}$-odd, which go like
\begin{eqnarray}
{\rm CP-odd, \hat{T}-even} &\propto& \sin\delta \sin\theta,\\
{\rm CP-odd, \hat{T}-odd} &\propto& \cos\delta \sin\theta.
\end{eqnarray}
The CP-odd and $\hat{T}$-even observables need a sizable
CP-conserving phase $\delta$ to show up, while the CP-odd and
$\hat{T}$-odd observables do not. On the other hand, in the presence
of the CP-conserving phase, the $\hat{T}$-odd observables may fake
the CP-violation effects since a CP-even and $\hat{T}$-odd goes like
$\sim \sin\delta \cos\theta.$ This indicates that caution must be
taken when studying only a $\hat{T}$-odd observable without the full
information of CP transformation for a state or a process.

There have been significant efforts in the literature to explore the
possibility to observe the effects of CP-violation from new physics
beyond the SM at the LHC. Examples include that in the top-quark
sector \cite{Schmidt:1992et,Atwood:1992vj,Brandenburg:1992be,topcpv}, and
in SUSY for a scalar top \cite{stop}. Most of the studies have been
concentrated on $\hat{T}$-odd observables \cite{topcpv,stop}, with
some of them also considering CP-odd observables in specific context
\cite{Schmidt:1992et,Atwood:1992vj,Brandenburg:1992be}.

In this Letter, we discuss a general approach to construct genuine
CP-odd observables at the LHC. We consider simple but common systems
that involve initial state partons. We present the general
discussion in section \ref{sec:general}, and show the illustrating
example in section \ref{sec:example}. The section
\ref{sec:conclusion} is devoted to the conclusion.

\section{CP-odd observables at the LHC}
\label{sec:general}

To construct genuine CP-odd observables at the LHC, we propose to study an
exclusive final state
\begin{equation}
f \bar f  + X^0,
\label{eq:fs}
\end{equation}
where $f$ is any charged particle(s) that can be kinematically reconstructed, $\bar f$ is its
charge conjugate, and $X^0$ is any particle(s) with neutral quantum numbers like charge, baryon
and lepton numbers.
With this event specification, the initial states must be from the
partons with neutral quantum numbers like $q\bar{q}$ or $gg$.

To avoid the ambiguity due to the longitudinal boost between the
partonic center-of-mass frame and the lab frame, we wish to seek for
kinematical observables involving only the quantities that are
invariant under the longitudinal boost, such as transverse
components of the momenta, and the direction of longitudinal momenta
difference. The simplest observables are the difference of the
transverse momenta, or equivalently the transverse energies,
\begin{equation}
\label{Teven}
p^+_T - p^-_T\quad  {\rm or}\quad  E^+_T - E^-_T,
\end{equation}
where $p^{}_T = \sqrt{p^2_x + p^2_y }$, $E^{}_T = \sqrt{ p^2_T + m_f^2 }$,
with $\pm$ specifying the charged particle.  This observable is CP-odd but $\hat{T}$-even.
Both CP-violating phase $\theta$ and CP-conserving phase $\delta$ are needed to generate
such observables. It would be sizable only if there is a large CP-conserving phase shift
in the final state interactions \cite{Schmidt:1992et}.

The next commonly used CP-odd variable is the triple product of the
three-momenta,
\begin{equation}
(\vec p_f\times \vec p_{\bar f}) \cdot \vec p_q .
\end{equation}
This is a $\hat T$-odd variable, and is generated by CP-violation in
the dispersive amplitude. Since the momentum direction of the
initial quark has the ambiguity with respect to which proton it is
from, this observable cannot be directly used at the LHC.
We thus consider a combination
\begin{equation}
\label{Todd}
(\vec p_f \times \vec p_{\bar f} ) \cdot \hat{p}_q\
{\rm sgn}( (\vec p_f - \vec p_{\bar f})\cdot \hat{p}_q).
\end{equation}
The factor  $(\vec p_f - \vec p_{\bar f})\cdot \hat{p}_q$ helps keep track
of the momentum directions, and it is the sign of it that is involved in
this  definition, which is invariant under a longitudinal boost.
The advantages of this variable are (1) it is a CP-odd and
$\hat T$-odd so that no CP-conserving phase is needed for it to be
generated; (2) it involves the quark momentum direction $\hat{p}_q$
twice so that it renders the specification of the direction
irrelevant. Therefore, we could even simply fix the direction of the
initial parton momentum along with a proton beam $\hat z$, without
changing its nature of transformation.

One could view this variable as a dot-product to define a polar
angle by $\cos\Theta$ of the $\vec p_f \times \vec p_{\bar f}$
vector with respect to the beam direction $\hat z$. CP violation
should thus manifest itself in the angular distribution
$\sigma(\cos\Theta)$. One may thus define a CP asymmetry
\begin{equation}
\label{eq:theta}
{\cal A}_{\Theta}^{CP} \equiv
\frac{\sigma({\cos\Theta>c_0}) - \sigma({\cos\Theta<-c_0}) }{\sigma({\cos\Theta>c_0})+\sigma({\cos\Theta<-c_0})},
\end{equation}
where $c_0$ is an appropriate selective cutoff for the asymmetry observation.
We find that, however, it is more intuitive to consider it as a
cross-product in the transverse plane $\vec p_{fT}\times \vec
p_{{\bar f}T} \sim (\vec p_f \times \vec p_{\bar f} ) \cdot
\hat{p}_q$, so that it defines an azimuthal angle $\Phi$ of $\vec
p_{{\bar{f}T}}$ with respect to $\vec p_{fT}$ in the range
$-180^\circ \le \Phi \le 180^\circ$.
%
One can now define an  asymmetry
\begin{equation}
\label{eq:phi} {\cal A}_{\Phi} \equiv
\frac{\sigma({\phi_1>\Phi>\phi_0}) - \sigma({-\phi_1<\Phi<-\phi_0})
}{\sigma({\phi_1>\Phi>\phi_0})+\sigma({-\phi_1<\Phi<-\phi_0})}.
\end{equation}
where, again,  $\phi_0$ is an appropriate selective cutoff for the asymmetry observation.

It is common that the CP asymmetries constructed above can take the
full angular range with $\phi_0=0^\circ, \phi_1=180^\circ$. It is
conceivable that a CP-violating interaction may lead to a shorter
period like in $\sin(n\Phi)$. Thus some caution needs to be taken in
constructing the asymmetry and in choosing the differential region
for $\phi_0$.
One could further extend the above discussion to include another factor
$\vec p_{fT}\cdot \vec p_{{\bar f}T} \sim \cos\Phi$. This will make the variable scale like
$\sin2\Phi$, and thus be more sensitive to the phase angle with a shorter period of $\pi$.
Keeping this factor may be particularly prudent given the unknown nature of CP-violation theories
and the potential complexity for the CP-violating operators.
%
The genuine  CP-odd and $\hat{T}$-odd observables  can be generalizations of that in Eq.~(\ref{Todd})
\begin{equation}
\label{eq:Todd} {\rm sgn}( (\vec{p}_f - \vec{p}_{\bar f})\cdot
\hat{z} ) \left( (\vec p_f\times \vec p_{\bar f}) \cdot \hat{z}
\right)^{2m+1} ( \vec p_{fT}\cdot \vec p_{{\bar f}T} )^{n} \sim
\sin^{2m+1}\Phi\ \cos^{n}\Phi\ ,
\end{equation}
with $m,n=0,1,2,...$.

It should be noted that we have chosen a very simple final state as
in Eq.~(\ref{eq:fs}). Along with the beam direction, there are only
three independent momenta, and thus only one triple cross product
can be constructed. It is desirable to generalize the situation to
include final states with more constructable particles, say $f_1\bar
f_1 + f_2 \bar f_2 + X^0$. Many more suitable CP-odd observables can
be obtained, depending on the underlying interactions and the
kinematics. The general construction principle discussed above
should still be valid if an initial state particle is involved in
the construction.

Let us summarize the points that have guided us for the construction
of genuine CP-odd observables for a system involving an initial
state parton at the LHC:
\begin{itemize}
\item[1.] Consider a set of particles that form a charge-neutral system;
\item[2.] Construct kinematical variables that are independent of the longitudinal boost
(involving only transverse components of their momenta);
\item[3.] For triple products, combine with the direction factor like ${\rm sgn}(\vec{p}_f - \vec{p}_{\bar f})\cdot \hat{z}$ to make the variable
insensitive to the choice of parton beam direction.
\end{itemize}
We expect that the above principles should be equally useful in constructing the genuine CP-odd observables for
new physics beyond the SM at the LHC:

\section{ Illustrating Example}
\label{sec:example}

We now demonstrate a genuine CP-odd variable discussed above by an explicit process at the LHC.
The simplest example of Eq.~(\ref{eq:fs}) could be $\ell^+\ell^- + \etmiss$.
We thus consider the $W^+W^-$ pair production at the LHC and their
subsequent leptonic decays \footnote{The CP-violation effect in this
process has been studied using $\hat{T}$-odd observable at the LHC
\cite{Kumar:2008ng} and CP-odd observables at the Tevatron
\cite{Dawson:1995wg}.}
\begin{equation}
pp \rightarrow W^+W^- \rightarrow \ell^+\ell^- \nu_\ell \bar{\nu}_\ell ,
\end{equation}
where we only consider $\ell=e,\mu$ for the sack of simple experimental identification.
In order to explore the effects of CP-violation, we adopt an
effective  lagrangian to parameterize the $WWZ$ interaction
\cite{Hagiwara:1986vm}
\begin{eqnarray}
{\cal L}_{WWZ}/g_{WWZ} &=& i
g^Z_1(W^{\dagger}_{\mu\nu}W^{\mu}Z^{\nu} -
W^{\dagger}_{\mu}Z_{\nu}W^{\mu\nu}) \\
&& + i \kappa_Z W^{\dagger}_{\mu}W_{\nu}Z^{\mu\nu}
+\frac{i\lambda_Z}{m^2_W}W^{\dagger}_{\lambda\mu}W^{\mu}_{\nu}Z^{\nu\lambda}
 -g^Z_4 W^{\dagger}_{\mu}W_{\nu}(\partial^{\mu}Z^{\nu}+\partial^{\nu}Z^{\mu})
\nonumber \\
&&+g^Z_5 \varepsilon^{\mu\nu\rho\sigma}(W^{\dagger}_{\mu}
(\overrightarrow{\partial_{\rho}}-\overleftarrow{\partial_{\rho}})W_{\nu})Z_{\sigma}
 +i \tilde{\kappa}_Z W^{\dagger}_{\mu}W_{\nu}\tilde{Z}^{\mu\nu}
+\frac{i\tilde{\lambda}_Z}{m^2_W}W^{\dagger}_{\lambda\mu}W^{\mu}_{\nu}\tilde{Z}^{\nu\lambda},
\nonumber
\end{eqnarray}
where
$\tilde{Z}_{\mu\nu}=\frac{1}{2}\varepsilon_{\mu\nu\rho\sigma}Z^{\rho\sigma}$.
In the SM, $g^Z_1=\kappa_Z=1$, and all the others are zero. We focus on
the CP-violating couplings, $g^Z_4$, $\tilde{\kappa}_Z$, and $\tilde{\lambda}$.
The most stringent  bounds on them are from an LEP-II experiment by DELPHI Collaboration
\cite{Abdallah:2008sf} as
\begin{eqnarray}
\nonumber
g^Z_4=-0.39^{+0.19}_{-0.20},\  \
\tilde{\kappa}_Z=-0.09^{+0.08}_{-0.05},\  \
\tilde{\lambda}_Z=-0.08\pm0.07.
\end{eqnarray}
We will use these non-zero central values in our further illustration.

\begin{figure}
\centerline{ \epsfig{file=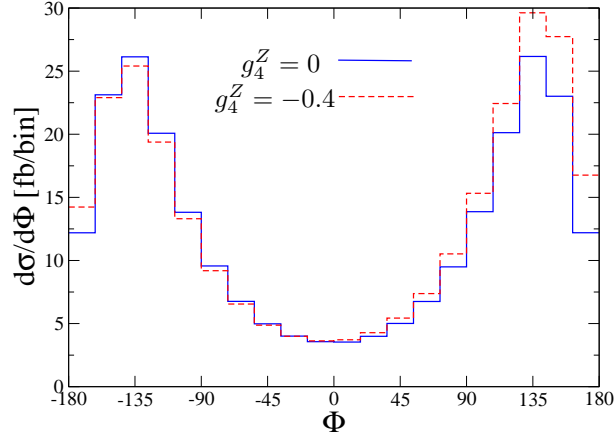,width=7.2cm,angle=-90}
\put(-165,-55){$g^Z_4=0$} \put(-175,-68){$g^Z_4=-0.4$} } \vskip
-0.5cm \centerline{\epsfig{file=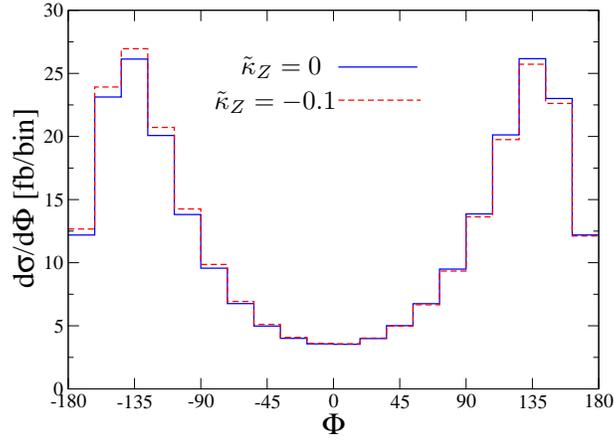,width=7.2cm,angle=-90}
\put(-165,-55){$\tilde{\kappa}_Z=0$}
\put(-175,-68){$\tilde{\kappa}_Z=-0.1$} } \vskip -0.5cm
\centerline{\epsfig{file=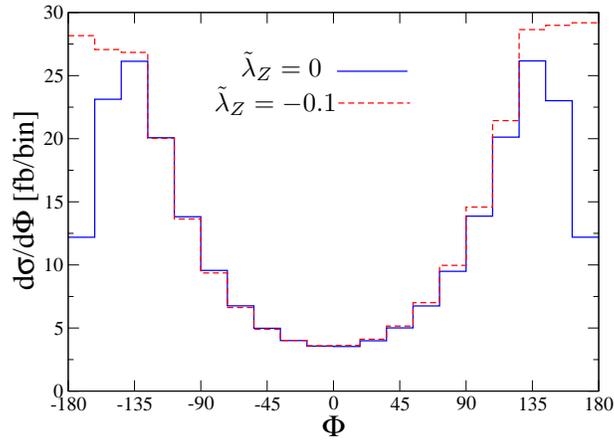,width=7.2cm,angle=-90}
\put(-165,-55){$\tilde{\lambda}_Z=0$}
\put(-175,-68){$\tilde{\lambda}_Z=-0.1$} } \caption{Differential
cross section with respect to the angle $\Phi$ in units of fb/bin ($18^\circ$)
for $pp\to W^+W^-\to \ell^+\ell^- \nu \bar\nu$ at the LHC (14 TeV),
with the CP-violating $WWZ$ couplings (a)
$g^Z_4=0,\ -0.4$, (b) $\tilde{\kappa}_Z=0,\ -0.1$, and (c)
$\tilde{\lambda}_Z=0,\ -0.1$ } \label{fig:dis_kap}.
\end{figure}

Our study is based on a Monte Carlo simulation of $W^+W^-$ pair production
and subsequent decays which incorporate the spin correlations. We use the
CTEQ6.1M parton distribution functions \cite{Pumplin:2002vw}. The
following cuts
\begin{eqnarray}
p_T(\ell) > 25\ {\rm GeV},\ \ \ptmiss > 25 {\rm GeV},\  \ |\eta(\ell)| < 2.5,
\end{eqnarray}
are implemented in our analysis \cite{cuts}. We smear the lepton
energy according to
\begin{equation}
\delta E/E = \frac{a}{\sqrt{E/{\rm GeV}}} \oplus b
\end{equation}
where $a=13.4\%$, $b=2\%$, and $\oplus$ denotes a sum in quadrature \cite{cuts}.

In practice, we can define an azimuthal angle
\begin{equation}
\Phi \equiv {\rm sgn}((\vec{\ell}^+ - \vec{\ell}^-)\cdot \hat{z})\ {\sin}^{-1} (\hat{\ell}^+ \times \hat{\ell}^-) \cdot \hat{z},
\end{equation}
with  $-180^\circ \le \Phi \le 180^\circ$. Here the particle names
$(\ell^\pm)$ have been used to denote their momenta.
One can now define the asymmetry
\begin{equation}
\label{eq:asymmetry} {\cal A}_{\Phi} \equiv
\frac{N_{\Phi>0}-N_{\Phi<0}}{N_{\Phi>0}+N_{\Phi<0}},
\end{equation}
where $N$ is the number of events.
A nonzero value of this asymmetry above the statistical error
$1/\sqrt{N_{total}}$ implies the CP-violation in this process.

In Fig.~\ref{fig:dis_kap}, we show the differential distributions for  the azimuthal angle $\Phi$
in the SM and with the anomalous couplings $g^Z_4$, $\tilde{\kappa}_Z$,
and $\tilde{\lambda}_Z$, in the units of fb/bin ($18^\circ$). With a 100 fb$^{-1}$
integrated luminosity, we expect a  large number of events, as indicated
by multiplying 100 on the left axis.
Significant asymmetry can be visible from the distributions.

\begin{table}[tb]
\caption{The number of events for $pp\to W^+W^-\to \ell^+\ell^- \nu \bar\nu \ (\ell=e,\mu)$
for $\Phi<0$ and $\Phi>0$ with and without the CP-violating couplings, at the LHC (14 TeV)
with a luminosity of 100 fb$^{-1}$. }
\begin{center}
\label{tab:eventnumber}
\begin{tabular}{|c|c|c|c|c|}
\hline  & $N_{\Phi<0}$ & $N_{\Phi>0}$ & $N_{\Phi>0}-N_{\Phi<0}$ & ${\cal A}_{\Phi}$ \\
\hline SM & $24700\pm160$ & $24700\pm160$ & $0\pm220$ & $\simeq 0.0\%$ \\
\hline $g^Z_4 = -0.4$ & $24600\pm160$ & $28600\pm160$ & $3950\pm230$ & $\simeq 7.4\%$ \\
\hline $\tilde{\kappa}_Z = -0.1$ & $25500\pm160$ & $24400\pm160$ & $-1130\pm220$ & $\simeq -2.3\%$ \\
\hline $\tilde{\lambda}_Z = -0.1$ & $28800\pm170$ & $30500\pm180$ & $1690\pm240$ & $\simeq 2.9\%$ \\
\hline
\end{tabular}
\end{center}
\end{table}

Integrating the differential cross-section over $\Phi$, we
obtains the asymmetry variable ${\cal A}_{\Phi}$. We find
\begin{equation}
{\cal A}_{\Phi}(g^Z_4 = -0.4) \simeq 7\% , \quad
{\cal A}_{\Phi}(\tilde{\kappa}_Z = -0.1) \simeq -2\%,\quad  {\rm and} \quad
{\cal A}_{\Phi}(\tilde{\lambda}_Z = -0.1) \simeq 3\%,
\end{equation}
respectively. With a
luminosity of 100 fb$^{-1}$ at the LHC, the number of events with
statistical errors are shown in table \ref{tab:eventnumber}. We
see about  17$\sigma$, 5$\sigma$, and 7$\sigma$,
signals of CP violation for cases with  $g^Z_4 = -0.4$, $\tilde{\kappa}_Z = -0.1$,
and $\tilde{\lambda}_Z = -0.1$, respectively.

The asymmetry in Eq.~(\ref{eq:asymmetry})  corresponds to the
observable in Eq.~(\ref{Todd}), which is proportional to ${\rm
sin}\Phi$. As discussed earlier, once the distribution of $\Phi$ is
obtained, one can construct  an asymmetry in any specific range
$(\Phi_0,\Phi_1)$ and $(-\Phi_0,-\Phi_1)$, as long as it is
statistically significant. More complex patterns of CP asymmetry (a
$\Phi$ distribution) may be counted for with the more general CP-odd
observables in Eq.~(\ref{eq:Todd}). For example, the term in
Eq.~(\ref{eq:Todd}) with $m=0,n=1$ is proportional to ${\rm
sin}\Phi{\rm cos}\Phi$. Thus a CP-odd distribution in $\Phi$ leads
to an asymmetry in the 1$^{st}$ and the  4$^{th}$, which is opposite
to that of the 2$^{nd}$ and the 3$^{rd}$) quadrats. Thus a
collective sum over $\Phi >0$ (in the 1$^{st}$  and the 2$^{nd}$)
and $\Phi >0$ (in the 3$^{st}$  and the 4$^{nd}$) would not lead to
an asymmetry due to a cancellation. In such circumstances, it is
necessary to consider the asymmetry of $\Phi$ for specific more
refined regions.

\section{Conclusion}
\label{sec:conclusion} We have discussed how to construct the
genuine CP-violating observables at the LHC experiments. We
considered simple but common systems that involve initial state
partons. We summarized our guiding principles at the end of section
II. We classified the observables according to the even and odd
properties under the naive T-transformation ($\hat{T}$). There are
two classes of observables of our interests: CP-odd and
$\hat{T}$-even, with the difference of transverse momenta as a
representative as in Eq.~(\ref{Teven}); CP-odd and $\hat{T}$-odd,
involving triple momentum product as in Eqs.~(\ref{Todd},
\ref{eq:Todd}). We expect them to have broad applications to many
processes in theories beyond Standard Model with CP violation. For
the purpose of illustration, we use simple example of $W^+W^-$
production and subsequent decays at the LHC, where the CP violation
effects are parameterized by effective CP-violating operators of
$WWZ$ couplings. We found that using the upper bounds of these
couplings allowed by the current collider experiments, the CP
asymmetries via these operators can be clearly visible with an
observable we constructed at the LHC with an integrated luminosity
of 100 fb$^{-1}$ at 14 TeV.

\vskip 0.2cm \noindent

{\it Acknowledgments:} T.H. was  supported in part by the
U.S.~Department of Energy under grant No.~DE-FG02-95ER40896.
Y.L.~was supported by the US DOE under contract DE-FG02-08ER41531.

\end{document}